\documentclass[aps, prd, onecolumn,showpacs,showkeys,amsmath,amssymb]{revtex4-1}
\usepackage{graphicx}  
\begin{document}

\title{Fundamental physics and cosmology with LISA}

\author{Stanislav Babak$^1$, Jonathan R. Gair$^2$, Antoine Petiteau$^1$, Alberto Sesana$^1$}

\address{(1) Albert Einstein Institute,  Am Muehlenberg 1, D-14476 Golm, Germany \\
(2) Institute of Astronomy, University of Cambridge, Cambridge, CB3 0HA, UK}
\email{stba@aei.mpg.de}

\begin{abstract}
In this article we briefly review some of the applications  to fundamental physics and cosmology of the observations that will be made with the future space-based gravitational wave detector LISA. This includes detection of gravitational wave bursts generated by cosmic strings, measurement of a stochastic gravitational wave background, mapping the spacetime around massive compact objects in galactic nuclei using extreme-mass-ratio inspirals and testing the predictions of General Relativity for the strong dynamical fields generated by inspiralling binaries. We give particular attention to some new results which demonstrated the capability of LISA to constrain cosmological parameters using observations of coalescing massive Black Hole binaries.
\end{abstract}

\maketitle

\section{Introduction}
\label{S:Introduction}
LISA is a proposed joint ESA-NASA mission which will be launched around 2025 and which will be sensitive to gravitational waves (GWs) of low-frequency (between $10^{-4}$ and $0.1$ Hz). Astrophysical sources that emit gravitational waves in this frequency range are significantly more numerous than those that emit in the high frequency range (above a few Hz) which is accessible to ground-based detectors. In our own Galaxy there will be $\sim 10^6-10^7$ ultra-compact white dwarf binaries which are emitting gravitational waves in the LISA band. The majority of these will not be individually resolved by LISA but will constitute a gravitational wave foreground below $\sim3$mHz \cite{Nelemans:2009hy, Nelemans:2009ea}. LISA will detect coalescing massive black hole (MBH) binaries throughout the Universe if the masses are in the right range ($\sim 10^4M_\odot$--$10^7M_\odot$). The parameters of these binaries will be determined to an unprecedented accuracy and the set of events LISA detects will provide valuable constraints on the processes driving galaxy formation and evolution~\cite{Gair:2010bx}.  At lower redshift, $z\lesssim 1$, LISA will observe extreme-mass-ratio inspirals (EMRIs), which follow from the capture of a compact stellar mass object (a white dwarf, neutron star or black hole) by a MBH from a surrounding cusp of stars in a galactic nucleus. These sources will offer an insight into the stellar dynamics and evolution of the central parsec of galaxies. All of this astrophysical information is very difficult or impossible to obtain in any other way but through GW observations with LISA \cite{Prince:2009uc}. 
  
Besides astrophysics, LISA will be a laboratory for fundamental physics, and it is on this application of LISA observations that we will focus on in this review. LISA will be able to detect 
gravitational wave bursts coming from cusps forming on cosmic strings if they exist. Strings are one dimensional topological defects that could be produced as a result of the breaking of a $U(1)$ symmetry. Such a symmetry breaking (and therefore cosmic string formation) is predicted by grand unification theory (GUT) \cite{stringsBook} and many string-inspired models of inflation~\cite{Sarangi:2002yt, HenryTye:2006uv}.  The low frequency operational band of LISA makes it more sensitive to these bursts than ground based GW detectors. 

In a similar way to the Cosmic Microwave Background, a stochastic gravitational wave background might have been formed in the early Universe as a result of the parametric amplification of  ``zero-point'' quantum oscillations \cite{Grishchuk}, during the first order phase transition 
during or at the end of inflation (\cite{Apreda:2001us, Kaminkowski1994, Baccigalupi1997}, see also \cite{Durrer:2010xc} and references therein) or during reheating
(\cite{Easther:2006vd, GarciaBellido:2007af, Dufaux:2007pt}, see also \cite{GarciaBellido:2010jp} and references therein). 
The present energy density of the stochastic  GW background is very uncertain, as it depends on poorly constrained parameters that enter models describing its production. However, if the 
density is above $\Omega_{GW} \approx 10^{-10}$ 
then it could be detected by LISA. Even if the GW background is strong enough, it will need to be distinguished from the GW foreground discussed above. This might be possible by using the cyclo-stationarity of the Galactic foreground, i.e., the fact that the level of the galactic foreground varies over a year as LISA becomes more or less sensitive to gravitational waves arriving from the direction of the galactic centre~\cite{Edlund:2005ye}.

Coalescing MBH binaries observed with LISA can be used as standard sirens (as opposed to the ``standard candles'' of electromagnetic astronomy) to probe the expansion history of the Universe~\cite{Schutz:1986gp, Holz:2005df}, see also \cite{VanDenBroeck:2010fp} and references therein. GW measurements provide an accurate estimate of the luminosity distance to a MBH merger. If an electromagnetic counterpart to a GW event can be identified, it will provide the redshift of the source and hence a point on the luminosity-distance/redshift relation that can be used to constrain cosmological parameters (like the cosmological constant, matter density, Hubble constant, spatial curvature etc.). In fact it is not necessary to assume that electromagnetic (e/m) identification of the host is possible. Instead, since LISA will observe about 30 events per year, it will be possible to use statistical techniques to derive constraints on the cosmological parameters without redshift measurements. This was the basis of the original standard siren proposal in~\cite{Schutz:1986gp}, and in this paper we will describe new results for LISA that appear here for the first time. In particular, we will assume  that all the cosmological parameters are known besides the effective equation of state of the dark energy and then focus on the precision with which LISA could measure that.

The study of stellar dynamics (especially the dynamics of the S-stars) in the Galactic center tells us that there is a very compact dark object with a mass $\sim 4\times 10^6 M_{\odot}$ in the Galactic nucleus 
\cite{Genzel:2010zy}. Observations and simulations lead us to expect the presence of a dark massive object (DMO) in the nuclei of almost all galaxies. The common assumption is that these DMOs will be mass black holes described by the Kerr metric, but we do not yet have direct observational evidence for the presence of an event horizon in these systems. I should be possible to use LISA observations of GWs generated during extreme-mass-ratio inspirals to directly probe the nature of the DMOs for the first time. A compact object captured by a DMO with mass in the LISA range 
will generate $\sim10^4-10^6$ cycles of gravitational radiation while it is orbiting in the strong gravitational field of the DMO. The emitted GWs encode an imprint of the spacetime surrounding the DMO and the hope is that we will be able to extract information from the GW signal which will allow us to test the nature of the DMO. The most promising approach described to date is to measure the multipole moments characterising the DMO (mass, spin and quadrupole moment) and compare them with the corresponding values for the Kerr metric.  

In this article we will review our current understanding of these topics. We will start, in Section~\ref{S:strings}, with a description of LISA's ability to detect GW bursts from cosmic strings before moving on to a discussion of how to detect a cosmological stochastic GW background in Section~\ref{S:stochastic}. In Section~\ref{S:cosmo} we will present new results on constraining cosmological parameters with LISA in the absence of electromagnetic counterparts using a statistical method. We  then review the possibilities for ``mapping spacetime'' using EMRIs in Section~\ref{S:mapping} before concluding the article with a summary in Section~\ref{S:summary}.

\section{Detecting gravitational wave bursts from cosmic strings}
\label{S:strings}
As mentioned above, a string soliton solution will arise in any gauge theory that has a broken $U(1)$ symmetry or in any phase transition in which $U(1)$ becomes broken during the evolution. 
Interest in cosmic strings originated in the context of grand unification theories (see \cite{stringsBook}  for a review on the subject), but it was later suggested that strings would also be produced by $D$-brane annihilation at the end of brane inflation~\cite{Sarangi:2002yt, HenryTye:2006uv, Tye:2002yb}. Some fraction of these strings will be in the form of finite size loops, and other strings will be infinite. The latter are particularly important as they stretch with the expansion of the Universe while the small loops quickly decay. When two strings collide they may reconnect with a probability $P_{rec}$ which is a property of the strings. This probability is almost unity for GUT-strings and has a range of values $[10^{-3}, 1]$ for brane-strings. The main parameter characterizing an individual string is its tension $\mu$. The two parameters ($\mu, P_{rec}$) characterize the properties of a network of strings
\cite{AllenShellard}. We refer readers to \cite{Polchinski:2004ia, Polchinski:2007qc} for nice reviews on the mechanisms for production of cosmic strings and to~\cite{CurtLISA9} for a review on prospects for their detection.

When a long string intersects itself, a closed loop will break off. If the size of the loop formed is smaller than the horizon, it will not expand and behaves like a massive object. The closed loops decay through emission of gravitational radiation and a gravitational wave background is produced from the incoherent superposition of all the signals generated by the decaying loops produced by the network over time. The spectrum of this  GW background is expected to be approximately flat over LISA's operational frequency band and can be characterized in terms of the energy density of the gravitational waves (GW) $\rho_{GW}$:
\begin{equation}
\Omega_{GW}(f) \equiv \frac{d\rho_{GW}/d \ln{f}}{\rho_c},
\end{equation}
where $\rho_c$ is the critical density of the Universe. This quantity depends on the tension of the strings as $\Omega_{GW} \sim \sqrt{G\mu}$ \cite{Polchinski:2007qc,  stringsBook, Polchinski:2004ia}. The proportionality coefficient depends on the scaling factor at the moment of loop formation $\alpha$ and on $\Gamma$, the scaling factor for the decay time of a loop (of length $l$) $t_d = l/\Gamma G \mu$.

At present, there is no observational evidence that cosmic strings exist, but assuming they do, current observations can set up an upper limit on the string tension. The most recent published limit on $\Omega_{GW}$ from the ground-based gravitational wave detectors, LIGO and Virgo, is $\Omega_{GW} (f\sim 100 Hz) \le 6.9\times 10^{-6}$~\cite{LVCNature}. A somewhat better limit has been derived from pulsar timing observations. A signature in the timing residuals of pulsar observations arises from the effect of propagation of electromagnetic pulses in a stochastic gravitational field. Analysis of current pulsar timing data has given an upper limit $\Omega_{GW}(f \sim 10^{-8} Hz)  \le 4\times 10^{-8} $, which corresponds to a limit on the tension of $G\mu < 1.5 \times 10^{-8}$ \cite{Jenet:2006sv, DePies:2007bm}. This limit will improve further as the observation time increases. However, the prospects with LISA are even better. It was shown in \cite{DePies:2007bm} that LISA would be able to detect a background from cosmic strings
with a tension $G\mu > 10^{-16}$.

When two long strings reconnect, two long kinked strings are produced. These kinks propagate along the strings and tend to straighten and diminish in strength over time as energy is emitted in gravitational wave radiation. Kinks thus produce short GW bursts, which could also potentially be observable by gravitational wave detectors. A second mechanism for the production of GW bursts from strings are cusps, which are created when a tiny part of the string propagates with a speed close 
to the speed of light. It was shown in  \cite{Damour:2001bk} that GW bursts generated by cusps will be more detectable  in GW observations than those generated by kinks, since the latter fall off more rapidly with frequency. The GW radiation from cusps is linearly polarized and highly beamed in the direction of propagation of the cusp. The shape of the burst in the time domain takes a very simple form, $h(t) \sim |t- t_c|^{1/3}$, where $t_c$ is the time of arrival of the burst at an observer. However, the cusp at $t=t_c$ exists only for an observer who lies \emph{exactly} along the direction of the cusp's velocity. For an observer in a direction at an angle $\theta$ to the center of the radiation cone, there is an exponential decay in the radiation spectrum at frequencies above $f_{max} \sim 1/(L\theta^3)$ where $L$ is a characteristic lengthscale of the cosmic string. This upper frequency cut off smoothes $h(t)$ at $t=t_c$~\cite{Damour:2001bk, Siemens:2003ra}. The GW waveform in the frequency domain scales as $\tilde{h}(f) \sim f^{-4/3}$ (as compared to the GW bursts from the kinks which scales as $\tilde{h}(f) \sim f^{-5/3}$). It was demonstrated in \cite{Cohen:2010xd} that LISA will be a factor of ten more sensitive to GWs from cusps than the advanced ground based detectors. The square of the signal-to-noise ratio (SNR) per logarithmic frequency interval ($d\ln{f}$) is approximately  $|f \tilde{h}(f)|^2/fS_h(f)$, where $S_h(f)$ is the noise  power spectral density. The SNR for a GW burst from a cusp therefore scales roughly as $f_b^{-1/3}/(f_b S_h(f_b))^{1/2}$, where $f_b$ is the frequency where the detector has the best sensitivity (this is $\sim 100$Hz for advLIGO and $\sim 0.003$Hz for LISA). The denominator ($(f_b S_h(f_b))^{1/2}$) for LISA is a few times higher than the corresponding value for advanced LIGO while the numerator, $f_b^{-1/3}$, is approximately 30 times larger for LISA. This means that, for a given SNR threshold, the volume in which LISA could detect bursts from cusps is a factor of $\sim 10^3$ larger than the observable volume for advLIGO. The majority of the bursts will be rather weak and will contribute to a gravitational wave background, however we would expect that a string network would also produce a few strong bursts which stand above the background. 

The detection of cosmic strings would have a great impact on fundamental physics and on our understanding of the early Universe. As mentioned above the string network can be characterized by two parameters $\mu, P_{rec}$, therefore if those two parameters can be determined we will learn about the nature of strings and about the network. An individual burst has no characteristic frequency scale and is characterized primarily by its amplitude, which is therefore not enough to determine the network parameters. However, it was argued by Polchinski \cite{Polchinski:2004ia} that the observation of $\gtrsim10$ bursts would provide the distribution of amplitudes, $dN \sim A h^{-B}dh$, and in principle the parameters $A$ and $B$ would constrain $\mu, P_{rec}$. However, any individual bursts that are observed will be at distances much closer than the Hubble distance and so the  burst distribution 
is likely to be close to the Euclidean, $dN/dh \sim h^{-4}$ \cite{Cohen:2010xd}, which does not constrain the network parameters. To place constraints on the string network, it may therefore be necessary to measure both the gravitational wave background ($\Omega_{GW}$) and the distribution of individual bursts $dN/dh$.

We will finish with a discussion of the detectability of bursts from cosmic string cusps using LISA. Since we know the form of the GW signal from a cosmic string cusp, we will be able to use matched filtering techniques to search for them in the LISA data stream. In order to assess existing data analysis techniques there have been a sequence of Mock LISA data challenges, and the last two of these have included bursts from cosmic string cusps in the data sets. These challenges involve blind searches for multiple signals in simulated LISA data \cite{Babak:2008sn, Babak:2009cj}.  For the cosmic string cusps,  the exact number of bursts in the data was not known (it was drawn from a Poisson distribution with mean 5).The injections used the following waveform model for the GW signal \cite{Babak:2009cj}:
\begin{eqnarray}
\tilde{h}_{ij}(f) = (A^{+}_{ij} + A^{\times}_{ij})A(f)e^{2\pi i ft_c},\\
A^{+}_{ij}  + A^{\times}_{ij} = \mathcal{A}\Lambda(f) (e^{+}_{ij}cos(\psi) + e^{\times}_{ij}sin(\psi))
\end{eqnarray}
\begin{equation}
\Lambda(f)  = \left\{\begin{array}{cc}
f^{-4/3} & f< f_{max} \\
f^{-4/3} e^{1-f/f_{max}} & f>f_{fmax}
\end{array}\right.
\end{equation}
in which the overall amplitude 
$
\mathcal{A} \sim G\mu l^{2/3}/D_L
$ is dependent on the luminosity distance to the source, $D_L$, the string tension and the characteristic scale of the string. The polarization angle $\psi$ and the polarization tensors are defined with respect to the fixed direction of GW propagation. 

The results of the blind search for the cosmic string bursts~\cite{Cohen:2010xd, Feroz:2009eb, Key:2008tt} (see also \cite{Babak:2009cj} for a summary of all the results for that round of the data challlenge) showed that current methods can be used to find GW bursts from cusps rather easily. Parameter estimation for these systems is a rather different issue, however, as the short duration of the signal means there are strong degeneracies in the parameter space. In particular the sky localization was rather poorly constrained (the error is more than one radian in both right ascension and declination) even if we are at the right maximum of the likelihood. Other parameters like the polarization angle and the amplitude strongly correlate with the sky location and were therefore also poorly determined. The parameter which will be determined most precisely is the time that the burst passes through the detector ($t_c$). Based on these results, we have very good prospects of detecting GW bursts from cosmic cusps with LISA but the parameters of the bursts will be rather poorly recovered.

\section{Stochastic gravitational wave background}
\label{S:stochastic}
There are different mechanisms which could lead to the production of a cosmological gravitational wave background. The primary mechanism is the adiabatic amplification of quantum fluctuations~\cite{Grishchuk}, which is most efficient during inflation. The resulting background is often referred to as the inflationary ``relic'' GW background. The other mechanisms involve classical (as opposed to quantum) sources and have the potential to be much stronger in certain frequency bands. One such mechanism is the decay of cosmic string loops, as we discussed in the previous section. There is a vast literature reviewing this topic \cite{Hogan:2006va, Maggiore:1999vm, Chongchitnan:2006pe, Buonanno:2003th,Hughes:2002yy} and so here we will just mention the various possible sources and the prospects for their detection with LISA.

A phase transition corresponding to symmetry breaking of the fundamental interactions could have happened in the early Universe. In the first order phase transition the Universe is initially trapped in a metastable  phase (with unbroken symmetries). The transition from the metastable phase to the ground state occurs by the quantum tunneling of a scalar field across a potential energy barrier. This transition nucleates randomly in bubbles. The size of these bubbles  increases as the temperature of the Universe drops and large bubbles then collide bringing the Universe to a broken symmetry phase. Gravitational radiation is produced as soon as the spherical symmetry of an 
individual bubble is broken during the collisions. The spectrum of GWs from the first order phase transition is peaked around a frequency determined by the typical temperature at which the transition takes place:
$f_{peak} \sim 100\, \rm{Hz} (T/10^5 TeV)$ \cite{Kosowsky:2000rq, Kaminkowski1994, Baccigalupi1997}. The electroweak phase transition happened at an energy scale of 100 GeV, for which the peak frequency is $\sim0.1$mHz, in LISA's sensitive frequency band. In the standard inflationary model, the electroweak transition is not of first order, although there are still Higgs field fluctuations (the vacuum expectation value undergoes a crossover from zero to a finite value), but these do not produce significant amounts of gravitational waves. If the standard inflation model is modified~\cite{Durrer:2010xc} (for example through the (next-to-)minimal supersymmetric extension of the standard model \cite{Apreda:2001us}) the electroweak phase transition could become strongly first order with a GW background of strength $\Omega_{GW} \le 10^{-12}-10^{-11}$ being generated through bubble nucleation. 

In addition, if the Reynolds number of the cosmic plasma is high and the growing bubbles convert internal energy into relativistic flows, this can generate high bulk turbulent velocities which accelerates the fluid leading to further production of GWs \cite{Kosowsky:2001xp, Gogoberidze:2007an, Durrer:2010xc}. The strength of the resulting GW background depends on the temperature of the Universe  at the time of the turbulence, the duration of this phase, the stirring scale and on the Reynolds and Mach numbers of the flows. If there is a small electromagnetic field (which should be present during the broken phase of the electroweak transition),  this field is amplified by the turbulence leading to a magnito-hydrodynamic turbulent plasma (MHD turbulence), which in turn generates further GWs.  Under favorable conditions the level of this background could be comparable or even higher than that coming from the bubble collisions and reach a level of $\Omega_{GW} \approx few \times 10^{-11}$~\cite{Durrer:2010xc}.

Other possible sources of stochastic GWs are reheating and the dynamics of extra dimensions. The presence of higher dimensions in, for example, brane-world scenarios, boosts the ratio of tensor-to-scalar perturbations in the early Universe and correspondingly increases the expected strength of the GW background. Inflation must end with a transition to a thermalized  Universe. This transition is called reheating, when the inflaton field decays into other degrees of freedom which also include the standard model particles. This process could generate significant relativistic inhomogeneities which would lead to the production of a GW background. However, the peak of the spectrum of such a background would lie at GHz frequencies for GUT-scale models of inflation. In the case of hybrid inflation, in which gauge fields associated with the breaking of some symmetries are present, the GW spectrum will have extra peaks associated with the mass scales of the corresponding gauge fields \cite{Dufaux:2010cf}. Unfortunately the predicted strength of this background is too weak to be observed by LISA, but it could be detected by BBO. This model also predicts an angular anisotropy of the GW background which potentially could be tested with the second generation of space-based GW observatories. If inflation ended with a global $O(N)$ transition, then the produced GW background would be at the level $\Omega_{GW} = 10^{-15} - 10^{-11}$~\cite{Fenu:2009qf}.  

Detecting a stochastic GW background and measuring its spectral shape will provide us information about the mechanism leading to its formation and the epoch in which this occurred, but how do we detect a GW background with LISA? In order to cancel laser phase noise, the six phase measurements at the spacecraft will be combined using a technique known as Time Delay Interferometry \cite{lrr-2005-4} to produce three TDI streams ($A,E,T$) in which the noise is uncorrelated (even for an unequal arm LISA \cite{Adams:2010vc}). The stochastic GW signal will be present in the $A$ and $E$ channels, but it will be strongly suppressed at low frequencies in the $T$ channel, which means a correlated  stochastic signal will be present in $A$ and $E$, while only instrumental noise will be present in $T$.  Using the correlation between $A$ and  $E$ and the ``null'' signal in $T$ at low frequencies allows the detection of a stochastic GW signal. The authors in \cite{Adams:2010vc} demonstrated using Bayesian techniques that LISA should be able to detect a GW background as weak as $\Omega_{GW} = 6\times 10^{-13}$. However, this estimate was obtained assuming instrumental noise only, i.e., the Galactic foreground from white-dwarf binaries was absent. The capability of LISA to detect a cosmological GW background with other sources in the data stream is yet to be fully quantified, although one early estimation~\cite{HoganBender2001} indicated $\Omega_{GW} \ge 10^{-11}- 10^{-10}$ as a detection threshold.

\section{Constraining cosmological parameters with LISA}
\label{S:cosmo}
The gravitational wave signals from coalescing MBH binaries that we expect to be present in the LISA data will, in general, be quite loud which will allow very precise measurements to be made of the source parameters. In particular, we expect to be able to determine the luminosity  distance of a binary  at redshift  $z=1$ to a  precision better than 1\% and to localize it on the sky to within 
$10 - 100$  arcminutes \cite{Lang:2008gh}. If the binary is embedded in a circumnuclear disk there might be a transient electromagnetic signal associated with the GW event and, for some nearby events, this e/m counterpart could be detectable (if the GW measurement provides sufficiently precise localisation of the source) which would provide a measurement of the redshift of the source in addition to the luminosity distance $D_L$ that can be measured from the GW data. Measurements of $D_L(z)$ tell us about the cosmological parameters describing the expansion history of the Universe. For this reason, coalescing MBH binaries are referred to as standard sirens~\cite{Schutz:1986gp,Holz:2005df, VanDenBroeck:2010fp} and much effort is currently being put into modelling the possible e/m signals which could accompany the merger of two MBHs (see for example~\cite{Schnittman:2010wy} and references therein).

In the following we will describe results that do not rely on making redshift measurements using e/m observations, but are instead based on a statistical approach to the problem. The motivation for 
this analysis is that e/m counterparts may be very weak or indeed not exist at all. We must then ask whether it is still possible to constrain the cosmological parameters by combining GW measurements from multiple sources. While the other sections of this article are reviewing previous work on the capabilities of LISA, the results in this section are new and appear here for the first time.

We can simulate the set of events that LISA will detect by using the model for the hierarchical formation of MBHs described in~\cite{Volonteri:2010py}. The model predicts the number of MBH binary merger events that LISA will see, and the SNR and MBH mass distribution of those events. We assume three years of LISA observations and generate many realizations of the set of events detected by LISA during its lifetime. Typically, we expect there to be about thirty events observed in three years of observation, up to a redshift of $z\le3$. We assume that the MBHs are spinning and that the spins point in random directions and have arbitrary magnitude. We adopt the model for the gravitational wave signal described in \cite{Petiteau:2010zu} which is a restricted waveform with the phase up to second post-Newtonian order. This model includes modulation of the amplitude and the phase due to  spin-orbital and spin-spin coupling but neglects all corrections in the amplitude (higher harmonics). 

\begin{figure}[ht]
\center{
\includegraphics[height=0.3\textheight,keepaspectratio=true]{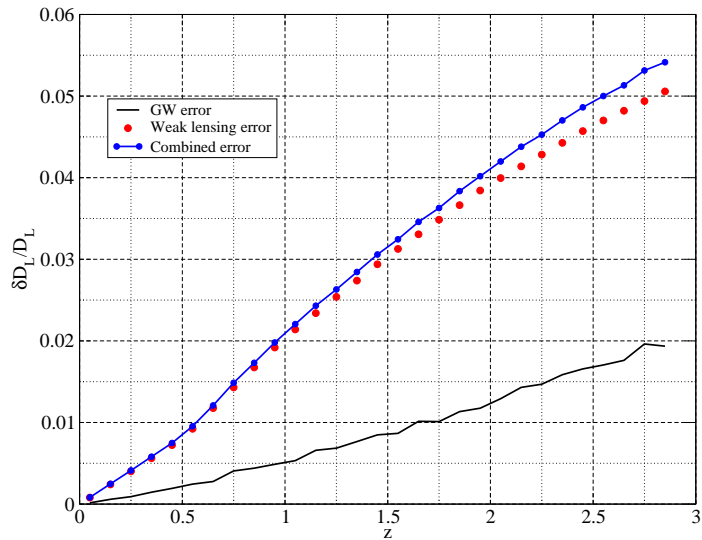}
}
\caption{Relative error in the luminosity distance due to instrumental noise (black solid line), 
due to weak lensing (red circles) and the combined error (blue circle-solid line). The weak lensing error 
is taken from \cite{Shapiro:2009sr}.}
\label{F:DLerr}
\end{figure}

We assume that the errors in the parameters determined from the GW observation have a multivariate Gaussian distribution with variance-covariance as predicted by the Fisher information matrix for this signal model. This is a reasonable assumption due to the expected high strength of the GW signals.  It was shown in~\cite{Petiteau:2010zu} that the likelihood surface for this waveform model contains several local maxima and some of these are comparable in likelihood to the global maximum. Here we will assume that the search algorithm used does manage to identify the true maximum. This will not be a trivial task, but we believe that it should be possible if we observe the signal for a reasonably long time (more than half a year) and observe the merger as well as the inspiral (some sky location degeneracies that exist at low frequencies could be broken as the GW signal propagates into frequencies above a few milli-Hertz).
  We are interested in the errors that are made in the estimation of  the luminosity distance and the sky location. To each gravitational wave event we may then associate an error box (a cylinder~\footnote{In fact, the error box should be an ellipsoid, but for simplicity we approximate it as a cylinder with 
  the cross-section equal to the $2\sigma$ error ellipse for the sky position and with the length equal to the $2\sigma$ error associated with the uncertainty in $D_L$}). The radius of the 
error cylinder is determined by the accuracy with which the GW source can be localized on the sky. The length of the cylinder along the line of sight is determined by the error associated with the measurement of $D_L$ and the uncertainties in the cosmological parameters, which are taken as priors. The error in $D_L$ arising from the GW measurement alone is quite small and dominates the error only at low redshift ($z<0.5$). At higher redshifts, the main source of distance error is weak lensing. Dark and bright matter along the line of sight to the source (de)magnifies the GW signal making it appear 
(further) closer than it is in reality. We model weak lensing using the mean errors quoted in~\cite{Shapiro:2009sr} and assume these are normally distributed, although strictly speaking this is not correct~\footnote{The fact
that the error in the luminosity distance associated with weak lensing is not gaussian could potentially 
improve our results. This will be explored in a follow-up publication.}. In Figure \ref{F:DLerr} we show 
the (median) error in the GW measurements due to instrumental noise as a black (solid) line, the errors due to weak lensing as (red) circles and the combined error as a (blue) circle-solid line.

The next ingredient required is a model for the Universe. We have used the Millennium simulation \cite{Lemson:2006ee} to represent the large scale structure of the Universe at different redshifts. The Millennium simulation assumes a lambda cold dark matter
model of the Universe and so do we. In particular, we model the evolution of the Hubble parameter as 
\begin{equation}
 H^2 = H_0^2 \left[ \Omega_m^0 (1+z)^3 + 
\Omega_{de}^0 \exp \left( 
3\int_0^z dz \frac{1+\omega(z)}{1+z}\right)  \right].
 \end{equation}
 with  $H_0 = 73.0\,\rm{km/(sec\, Mpc)},\; \Omega_m^0 = 0.25, \;\Omega^0_{de} = 
0.75,\; \omega(z) = -1$.
Using the semi-analytic model described in \cite{De_Lucia:2006vua}, one can attach to each dark-matter halo the masses of the components of each galaxy and their luminosity.  
Galaxy evolution is followed `self-consistently' by including in each protohalo identified in the simulation at high redshift a $10\%$ mass fraction in baryons, in the form of cooling gas. Baryon evolution is then followed taking into account a whole series of physical processes including cooling, star formation, heating by supernova feedback, bulge formation triggered by mergers, secular disk formation, etc. The parameters defining the recipes describing each of the physical processes are tuned in order to reproduce a broad range of observations (galaxy luminosity functions at different redshift, color distributions, etc)~\cite{DeLucia:2005yk}.

For each GW event in our realisation of the LISA data, we choose the nearest snap shot of the Millennium simulation and associate one of the directions in the comoving volume as the direction of the line of sight (redshift). There are a total of 63 snapshots which are logarithmically spaced in redshift. We have used snapshots corresponding to the redshifts $\{z = 0.51, 1.08, 1.50, 2.07, 2.62, 3.06\}$. Then we chose the galaxy that is the host of the merging MBHs. The probability of the galaxy to be a host is taken to be proportional to the local density. We note, however, that the probability that the host is in a low density region is not small (about 50\%), since although the probability that the merger happens in a low density part of the universe is low, there are many such parts with low local density.  Once the host is chosen it is surrounded by the error box (cylinder) described above. In practice we need to place an error box in the sky coordinates and in the  redshift and in order to go from the measured luminosity distance to the redshift we must assume a cosmological model. As a result the size of the error box is determined not only by the errors associated with the determination of the luminosity distance but also by the cosmological uncertainty in translating the $D_L$ to a redshift.  We assume that all cosmological parameters are known except for the effective equation of state of dark energy, which is modelled as $\omega(z) = -1 + w$. We use the currently estimated range for $w$, which is $w \in [-0.3:0.3]$.

 Our main objective is to show that by using all GW observations together it is possible to tighten these constraints on $w$. A similar statistical approach was 
employed in \cite{MacLeod:2007jd} to reduce the uncertainty in $H_0$ as derived from GW observations of extreme-mass-ratio inspirals with LISA. The basic idea is to measure the redshift of all galaxies which are potential hosts of the merger, i.e., which lie within the error box, combine these statistically into an estimate of the source redshift, and correlate the results across all GW events. We emphasize several important points:
(i) we assume that, by the time LISA flies, we will be able to measure the redshift to all galaxies with an apparent magnitude $m \le 24$, which means there is a strong selection effect at high redshifts, at which we will see only very massive galaxies, whereas LISA mergers are most likely to be associated with 
moderate mass $\sim 10^{10} M_{\odot}$ merging galaxies. The consequence of this assumption is that the host may not be included in the set of galaxies used, but this was also true in~\cite{MacLeod:2007jd} and it was seen not to be a problem in that case. We investigate the importance of this selection effect for the current work in a separate publication~\cite{Babak_inPrep2010}. (ii) The method relies on the fact that the distribution density within the error box is not uniform. The larger the contrast in density along the line of sight the better the method works. For this reason we do not consider GW events at redshifts $z\ge 3$ (iii) Since we might not be able to observe hosts at moderate to high redshifts, we rely on the similarity in the density profile across different mass ranges, i.e., that the spacial distribution of the lighter galaxies which host LISA sources traces the distribution of the more massive galaxies that we can observe. We verified the similarity of the different mass distributions using Millennium data. In our simulations we choose the host from all galaxies in the snapshot but for measuring the redshift we use only the galaxies with apparent magnitude $m\le 24$. We also use only the bright galaxies to estimate the local density. If the chosen host is close to the boundary of the snap shot box, we assume periodic boundary conditions for the data (as was done in the Millennium simulation itself). 

\begin{figure}[ht]
\center{
\includegraphics[height=0.25\textheight,keepaspectratio=true]{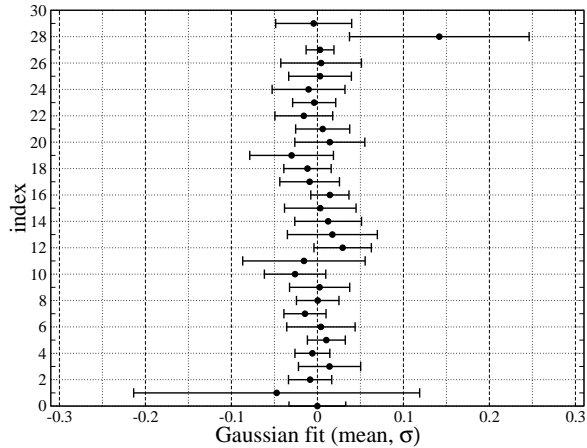}
}
\caption{Parameters of the gaussian fit for the final probability density $P(w)$ for each of the 30 realizations.}
\label{F:summary}
\end{figure}

The posterior (marginalized) density distribution for $w$ from an observation ($s$) of a single GW event is 
\begin{eqnarray}
P_j(w | s) &=& \frac{p_0(w) P_j(s | w)}{E_j} , \\
P_j(s | w) &=& \int \Lambda_j(D_L(z,w), \theta, \phi) p_j(\theta, \phi, z)\; d\theta\; d\phi\;  dz; \;\;\;\;
E_j = \int p_0(w) P_j(s | w) dw
\label{poster}
\end{eqnarray}
where $p_0(w)$ is the prior on $w$, $\Lambda_j(D_L(z,w), \theta, \phi)$ is the likelihood 
marginalized over the other parameters of the MBH binary so it depends only on $D_L$ and the sky
position ($\theta, \phi$), and  $p_j(\theta, \phi, z)$ is the astrophysical prior for a given galaxy to be the host, which is proportional to the square of the local density. The final distribution for $w$ (assuming that the individual events are independent) is constructed from the products of the individual $P_j(s|w)$: 
\begin{equation}
P(w) = \frac{p_0(w) \prod_j P_j(s| w)} {\int p_0(w) \prod_j P_j(s | w) dw}. 
\end{equation}
We considered 30 realisations of the LISA data and used a flat prior on $w$. The final probability density 
$P(w)$ could be well approximated in most of the realisations by a single gaussian and in all other cases by a sum of a small number of gaussian profiles. The peak of the distribution was usually found to be close to zero (which was the true value of $w$ used to generate the data set) and the ($1\sigma$)width of the gaussian was a factor 2-8 smaller than the prior range. A summary of the results of the simulations is shown in Figure~\ref{F:summary}.
To produce this figure, we have fit the posterior in each realisation using a Gaussian and the figure shows the mean (as a circle) and the variance (as an error bar) of the fitted Gaussian for each realisation. In a few cases there was a nearby merger which could be very well localized on the sky, and there was then only one cluster of galaxies in the  error box, which is essentially equivalent to having an e/m counterpart. The GW events for which we could identify  the e/m counterpart in this way were removed from the analysis to allow us to show how well we can constrain $w$ without an e/m identification of the host. The figure shows that we have very good prospects for constraining the effective equation of state of dark energy provided that we know all the other cosmological parameters (which could be the case by the time LISA is launched). In a separate publication~\cite{Babak_inPrep2010} we will present a more detailed description of our simulations and an analysis of a larger number of realizations with varying parameters (including the effect of using a different weak lensing model or a deeper spectroscopic survey, the effect of including or omitting events with an identified e/m counterpart in the cases where these are present, etc).

\section{Mapping spacetime around compact massive objects in galactic nuclei}
\label{S:mapping}
In this final section of the review, we turn our attention to probing the nature of the DMOs that are observed to be present in galactic nuclei. The common belief is that these DMOs are Kerr BHs, but can we test this assumption? One of the ways to answer this question is through observations, with LISA, of GWs from extreme-mass-ratio inspirals. In such systems, the inspiralling compact object (CO) can spend a significant amount of time in the strong field region of spacetime close to the central object before it plunges. Detailed information about the structure of the spacetime that the CO is orbiting in will be encoded in the GW signals that are received by LISA. If a strong EMRI signal is detected, then we will first need to extract this information, and, if a deviation from ``Kerriness'' is detected, we will need to interpret it. We refer readers to~\cite{Drasco:2006ws} for a review on EMRI waveforms and to~\cite{AmaroSeoane:2007aw} for a review of related astrophysics and concentrate here only on the use of the EMRI signals for mapping spacetime.

The question is, what kind of information in a GW signal would indicate a deviation from the Kerr spacetime in the host system? GW measurements with LISA will rely on matched filtering and will therefore be sensitive to effects that lead to a mismatch in the phase between the signal and the model. For EMRIs, we will detect GWs over $10^4-10^6$ orbits of the CO, and so we will be sensitive to relatively small changes in the orbital motion. Most of the research on this question to date has focussed on how the orbits of test particles differ in spacetimes that deviate from Kerr. We review below the various attempts that have been made to construct and characterize such spacetimes.

The first approach to spacetime mapping was proposed in~\cite{Ryan:1995wh}. This relied on using the multipole decomposition of the metric outside a stationary and axisymmetric central body in general relativity, which can be expressed in terms of mass, $M_l$, and current, $S_l$, multipole moments. For a Kerr BH these moments are determined by two parameters --- $M_0 = M_{BH}$, the total mass of the BH and $S_1 = M_0 a$, the spin --- and all higher moments are determined from these through the relation $M_l + iS_l = M_{BH} (ia)^l$. If more than two moments of a spacetime were measured using GW observations, these could then be tested for consistency with the Kerr solution. It was shown that the different multipole moments enter at different orders in a decomposition of the vertical and radial epicyclic frequencies as a function of orbital frequency for circular-equatorial orbits~\cite{Ryan:1995wh}. These epicyclic frequencies can, in principle, be determined from a gravitational wave observation, allowing the multipole moments to be measured. However, the higher multipoles enter the phasing of the GWs waves with a strength that decreases with multipole number, so in practice we would expect only to be able to measure the lowest few moments. Nonetheless, for an EMRI with masses $10\, M_{\odot}, 10^6\, M_{\odot}$ and with SNR $\sim 100$ it is estimated that we will be able to measure the quadrupole moment of the central object with a precision of $\sim10^{-3}$~\cite{Barack:2006pq}, while simultaneously measuring the mass and spin of the black hole to $\sim10^{-4}$.

The multipole decomposition is not a convenient way to characterize nearly-Kerr spacetimes, since the Kerr spacetime has an infinite number of non-zero multipoles. For that reason, a number of authors have considered various ``bumpy'' BHs, which are solutions to the Einstein field equations that are close to Kerr but depend on a deviation parameter, $\epsilon$, such that for $\epsilon=0$ the spacetime reduces to the Kerr metric. This deviation parameter can in principle be measured or bounded by a GW observation.
\begin{itemize} \item The first such analysis was described in~\cite{Collins:2004ex}. They constructed a bumpy BH using a perturbation of the Schwarzschild metric. This was later extended to rotating objects using an elegant new algorithm in~\cite{Vigeland:2009pr}. In these papers the authors analyzed how the deviations from the Kerr solution modified the fundamental orbital frequencies of the geodesics in the spacetime. 
\item A different approach was taken in \cite{Glampedakis:2005cf} in which the bumpy BH was inspired by the Hartle-Thorne metric for slowly rotating bodies. A kludge waveform (without radiation reaction) was constructed and for the first time the confusion problem was considered, i.e., that there could exist a Kerr waveform which is a near-perfect match to the waveform of a quasi-Kerr spacetime with slightly different parameters, if radiation reaction is not taken into account. This ``quasi-Kerr'' metric was used in \cite{Johannsen:2010xs} and follow up papers to analyze the propagation of null geodesics and to consider the imprint of non-Kerrness on electromagnetic observations. 
\item In~\cite{Gair:2007kr} the authors considered an exact solution in GR (Manko-Novikov) describing an arbitrary axisymmetric body which can deviate from Kerr at any multipole moment. They again studied the frequencies of geodesic motion and the existence of a third integral of the motion (which is the Carter constant in the Kerr spacetime) as well as precession rates in this general spacetime. It was found that the third integral could be lost under certain circumstances, leading to ergodic motion, which would be a ``smoking-gun'' for a deviation from Kerr if it were observed. 
\end{itemize}
In each of the above papers, the main observables considered were the frequencies of geodesic orbits, deviations in which which would show up in GW observations through small phase-shifts over a long observation. The loss of the third integral considered in~\cite{Gair:2007kr} is a more robust signature of a deviation from the Kerr metric as that behaviour is qualitatively different from what is expected in the Kerr spacetime. Another example of a feature that is qualitatively different in a perturbed spacetime was suggested in \cite{LukesGerakopoulos:2010rc}. When an integrable system is perturbed, resonant points become smeared out into resonant chains of islands (the Poincar\'{e}-Birkhoff theorem). Such perturbations would manifest themselves as a persistent resonance in the observed GWs, i.e., by a persistent period of time in which two of the fundamental frequencies of the orbit were commensurate.

In all the above cases, the authors dealt with vacuum solutions of GR, but it is clear that astrophysical black holes will not be pure vacuum solutions. There have also been several attempts to describe a deviation from Kerr due to the presence of matter, for instance by a scalar field with considerable self-interaction (a so-called massive Boson star)~\cite{Kesden:2004qx}. The non-rotating Boson star has no horizon and is not as compact as a BH, so the orbit of an inspiralling compact object could pass inside the Boson star, where the metric deviates from Schwarzschild. This will leave an imprint on the GW signal, as the GW emission will persist after the plunge should have been observed. Another possibility is that the Kerr BH is surrounded by an accretion disk. The authors in~\cite{Barausse:2006vt} considered the gravitational influence of a massive self-gravitating torus around a Kerr MBH on the orbit of a CO that did not intersect the disc. If radiation reaction is not taken into account, one can readjust the orbital parameters, e.g., the mass and spin of the MBH, in the presence of the torus so that the confusion problem (i.e., that we cannot distinguish the waveform from a pure Kerr waveform) discussed earlier is seen. The amount of material required to leave an imprint on the signal was found to be unreasonably high, suggesting that the gravitational influence of an accretion disc would be undetectable even when radiation reaction was taken into account. However, in a follow up paper~\cite{Barausse:2007dy}, the effect of the hydrodynamical drag force on the particle when the CO orbit intersected the disc  was considered. In that case, a small amount of energy would be dissipated due to the interaction of the body with the gas on each passage through the disc (possibly several times per orbit). This interaction would lead to small changes in the orbital radius and eccentricity and to a decrease in the orbital inclination, which is qualitatively different to the pure-GR case in which radiation-reaction drives an increasing inclination. However, for the drag force effect to be comparable to the radiation reaction force (self-force) one would again need a very massive accretion disk and a relatively low-mass MBH ($\sim 10^5\, M_{\odot}$). Recently it was shown \cite{Yunes:2010sm} that the presence of a second MBH within a few tenths of a parsec could also leave a measurable imprint on the EMRI waveform. In that case, the effect arises from the fact that the center of mass of the EMRI system is no longer an inertial frame due to the acceleration exerted by the perturber which this leads to a Doppler phase shift of the GW signal.

In all of the proceeding analyses, the calculations were done within the theory of GR. However, deviations in EMRI observations might also arise if the true theory of gravity is \emph{not} GR. In \cite{Sopuerta:2009iy}, the authors considered BHs in Chern-Simons theory, which deviate from Kerr BHs in the fourth multipole moment. This affects geodesic motion and correspondingly the phase of the GW signal. The authors found that the deviations in geodesic orbits could be significant, but more work is required to estimate the precision with which EMRI observations will be able to constrain a Chern-Simons modification to GR. EMRI observations can also be used to test theories of massive gravity. In these theories the GWs have additional polarizations~\cite{Babak:2002uz}, and propagate with a speed different from the speed of light. One would thus expect to see a dispersion when comparing different harmonics of an EMRI signal. If we are lucky enough to see a disruption of a WD by a $\sim 10^5\, M_{\odot}$ MBH \cite{Sesana:2008zc}, then we could also compare the time delay between the GW and electromagnetic signals. EMRIs can also be used to constrain scalar-tensor theories, but this requires the inspiralling object to be a neutron star rather than a black hole. In \cite{Berti:2005qd} it was estimated that observations of a neutron star inspiralling into a $10^4\, M_{\odot}$ spinning BH with SNR $\sim 10$ could bound the Brans-Dicke parameter, $\omega_{BD}$, to $\omega_{BD}>$few$\times 10^3$, which is slightly better than current Solar System bounds from observations of the Cassini satellite.

While most research to date on testing GR using LISA has focussed on EMRIs, test will also be possible using observations of coalescing MBH binaries of comparable mass. In such systems, the deviation from GR will again leave an imprint on the inspiral and correspondingly on the phase of the GWs. For instance, in \cite{Berti:2005qd} the authors estimated the precision with which effects coming from massive gravity could be measured with LISA. They showed that by observing a spinning $M \sim 10^6\, M_{\odot}$ comparable mass MBH binary LISA would be able to constrain the Compton wavelength of the graviton down to $\sim 10^{-16}$ (in units $M$). In addition, the post merger GW signal (quasi normal mode ring-down) depends on the final MBH which, in GR, is characterized only by its mass and spin. If we were able to detect two or more harmonics of the ringdown waveform, we would be able to test the no-hair theorem by making a consistency check~\cite{Berti:2005ys} that the mass and spin measured from each of the harmonics were consistent. This test will only be practical for comparable-mass systems, as the amplitude of ringdown radiation after an EMRI will be too small to allow detection.

\section{Summary}
\label{S:summary}
In this article we have discussed the capability of LISA as a laboratory for testing fundamental 
physics. LISA, operating in a low GW frequency band, will be sensitive enough to detect or to set 
upper bounds on cosmic string networks and the stochastic gravitational wave background that are much better than those currently available. By analyzing the phasing of the GWs emitted during extreme-mass-ratio inspirals and the coalescence of MBHs we will be able to extract information about the deviation of the spacetime in the host systems from the pure (vaccum) Kerr solution of relativity. Possible sources of such deviations include the presence of a non-Kerr object in the spacetime, perturbing material around the black hole or a deviation of the theory of gravity from general relativity. 

We have also presented new results on the capability of LISA to constrain cosmological parameters through GW observations of MBH binaries out to high redshift ($z\le3$). In particular, if  we  assume that the only unknown parameter is the effective equation of state for the dark energy, $w$, then, using a statistical method, if LISA observes approximately 30 MBH binaries we will be able to reduce the current uncertainty $|\delta w | < 0.3$ by a factor of 2 to 8.

\begin{acknowledgments}
S.B. would like to thank Curt Cutler for help in preparing section on cosmic strings. 
\end{acknowledgments}

\section*{References}
\bibliographystyle{apsrev4-1}

\bibliography{Refs}


\end{document}